\documentclass[a4paper]{article}
\usepackage{spconf,amsmath,graphicx,amssymb,delarray,subfigure,verbatim,booktabs}
\usepackage[colorlinks,linkcolor=black]{hyperref}
\usepackage{multirow}
\usepackage[english]{babel}

\title{DYNAMIC ATTENTION BASED GENERATIVE ADVERSARIAL NETWORK WITH PHASE POST-PROCESSING FOR SPEECH ENHANCEMENT}
\name{Andong Li$^{\star \ddagger}$, Chengshi Zheng$^{\star \ddagger}$, Renhua Peng$^{\star \ddagger}$, Cunhang Fan$^{\dagger \ddagger}$, Xiaodong Li$^{\star \ddagger}$}
\address{$^{\star}$ Key Laboratory of Noise and Vibration Research, Institute of Acoustics, Chinese Academy\\
	of Sciences, Beijing, China\\
	$^{\dagger}$ NLPR, Institute of Automation, Chinese Academy of Sciences, Beijing, China\\
	$^{\ddagger}$ University of Chinese Academy of Sciences, Beijing, China}

\begin{document}
	\ninept
	\hyphenpenalty=8000
	\tolerance=800
	\maketitle
	\begin{abstract}
	The generative adversarial networks (GANs) have facilitated the development of speech enhancement recently. Nevertheless, the performance advantage is still limited when compared with state-of-the-art models. In this paper, we propose a powerful \textbf{D}ynamic \textbf{A}ttention \textbf{R}ecursive GAN called DARGAN for noise reduction in the time-frequency domain. Different from previous works, we have several innovations. First, recursive learning, an iterative training protocol, is used in the generator, which consists of multiple steps. By reusing the network in each step, the noise components are progressively reduced in a step-wise manner. Second, the dynamic attention mechanism is deployed, which helps to re-adjust the feature distribution in the noise reduction module. Third, we exploit the deep Griffin-Lim algorithm as the module for phase post-processing, which facilitates further improvement in speech quality. Experimental results on Voice Bank corpus show that the proposed GAN achieves state-of-the-art performance than previous GAN- and non-GAN-based models.
	\end{abstract}

	\section{Introduction}
	\label{sec:intro}
	Speech enhancement (SE) is regarded as the technique to extract the speech components from the noisy signals, which helps to improve the speech quality and speech intelligibility~{\cite{loizou2013speech}}. It is widely used for automatic speech recognition (ASR), hearing assistive devices and speech communication. Recently, due to the tremendous capability of deep neural networks (DNNs) in modeling complicated non-linear mapping functions, a multitude of DNN-based SE approaches have been proposed to recover the speech components in low signal-to-noise ratio (SNR) and unstable noise environments~{\cite{wang2018supervised, tan2019learning}}. This paper focuses on monaural speech enhancement task. 
	
	Recently, generative adversarial networks (GANs) have been widely used in image-to-image translation tasks~{\cite{schonfeld2020u}}, which receive considerable attention from the speech research community~{\cite{pascual2017segan, kaneko2017generative, michelsanti2017conditional}}. They encompass two principal parts, namely generator network (G) and discriminator network (D), which are optimized by playing a min-max game between each other~{\cite{goodfellow2014generative}}. The objective of the generator is to synthesize the fake samples which resemble the target data distribution while the discriminator attempts to discriminate between the real and fake samples. SEGAN is the first network incorporating GAN for SE task, where the speech is enhanced directly in the time domain~{\cite{pascual2017segan}}. Nonetheless, no notable improvement in objective metrics is observed than the traditional signal-processing approach. Afterward, more training strategies are introduced, which facilitate better performance for time-domain based GANs~{\cite{fan2019noise, baby2019sergan, liu2020cp}}. Another line of research is based on the time-frequency (T-F) domain, where G is to map the noisy T-F features to the corresponding T-F targets~{\cite{soni2018time}}. The experimental results indicate that T-F masking-based approaches are more beneficial for noise reduction than SEGAN~{\cite{soni2018time}}.
	
	Despite the fact that impressive performance has been achieved for various GAN-based SE approaches, they still have several drawbacks, which are attributed as three-fold. First, although the time-domain-based GANs effectively circumvent the phase estimation problem, they make it more challenging to optimize D. This is because waveform has less structural characteristics than T-F representation. For example, the frequency information is implicitly determined through neighboring points in the time domain whilst the frequency distribution is explicitly represented when transformed into the T-F domain. As a consequence, D has a better discriminative capability in the T-F domain. Second, most of the networks adopt complicated topology for better performance. However, Ren \textit{et al.} proposed a simple baseline by unfolding the shallow network repeatedly, which achieved state-of-the-art (SOTA) performance in the  deraining task~{\cite{ren2019progressive}}. It reveals the significance of the multi-stage training protocol. Third, most T-F domain-based GANs estimate the magnitude of the spectrum, leaving the phase information unprocessed, which causes the phase mismatches.
	
	Motivated by our proposed dynamic attention recursive convolutional network (DARCN)~{\cite{li2020recursive}}, we propose a novel GAN-based model noted as DARGAN. Compared with previous GANs, innovations can be summarized as three-fold. First, a recursive protocol is utilized during the training of G, i.e., different from directly generating the fake samples in G, the mapping procedure is decomposed into multiple stages, where the dependencies across the stages are bridged through a memory mechanism. Second, a dynamic attention mechanism is introduced, where the attention generator network is specifically designed to control the feature distribution of noise reduction network. Third, different from directly reconstructing the waveform with noisy phase information, we introduce a type of phase post-processing (PPP) technique based on deep Griffin-Lim algorithm (DGLA)~{\cite{masuyama2019deep}}, which is capable of effectively reconstructing the phase information via unfolding the block for multiple times. Experimental results indicate that the proposed model achieves SOTA performance than previous GAN-based SE models.
	
	\begin{figure*}[t]
		\vspace{-0.3cm}
		\setlength{\abovecaptionskip}{0.235cm}
		\setlength{\belowcaptionskip}{-0.1cm}
		\centering
		\centerline{\includegraphics[width= 1.8\columnwidth]{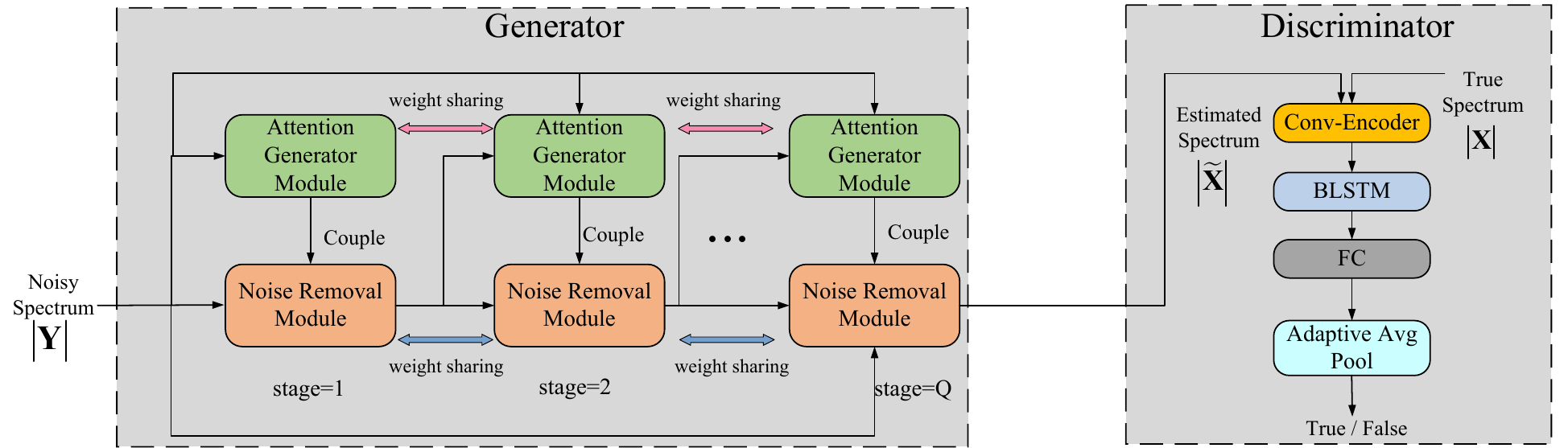}}
		\caption{The schematic of proposed architecture. It consists of two parts: a generator network (G) and a discriminator network (D). G comprises two modules, namely Noise Removal Module (NRM) and Attention Generator Module (AGM). $Q$ denotes the number of stages in G.}
		\label{fig:proposed-architecture}
		\vspace{-0.5cm}
	\end{figure*}
	
	The remainder of the paper is organized as follows. The concept of GAN is introduced in Section~{\ref{sec:generative-adversarial-network}}. Section~{\ref{sec:proposed-network}} explains the proposed network. Section~{\ref{sec:experimental-settings} gives the experimental settings. Results and analysis are illustrated in Section~{\ref{sec:results-and-analysis}}. We draw some conclusions in Section~{\ref{sec:conclusion}}.
	
	\vspace{-0.2cm}	
	
	\section{Generative Adversarial Network}
	\label{sec:generative-adversarial-network}
	\vspace{-0.2cm}	
	Generative adversarial network (GAN) is first proposed by Goodfellow et. al.~{\cite{goodfellow2014generative}}. It is comprised of two parts, namely generator network (G) and discriminator network (D). G aims to map the noise variable $\mathbf{z}$ from the prior distribution $p_{\mathbf{z}} \left( \mathbf{z} \right)$ to generated fake samples $G\left(\mathbf{z}; \theta_{g} \right)$. As for D, it is trained to accurately recognize whether the input is from the generated samples (fake) or training data $p_{data} \left( \mathbf{x} \right)$ (real). Both components are optimized by playing a min-max game.
	
	In GAN-based SE models, conditional GAN (cGAN) is usually adopted~{\cite{isola2017image}}, i.e., G generates the enhanced speech based on the prior noisy speech input. In this study, the waveform is first transformed into T-F domain with the short-time Fourier transform (STFT), and then the amplitude spectrum is utilized as both the input and output during the sample generating process. Least-squares GAN (LS-GAN)~{\cite{mao2017least}} is employed for training. Assuming the input noisy features, clean targets, and the generated version are notated as $| \mathbf{Y} |$,  $| \mathbf{X} |$, and $| \mathbf{\tilde{X}} |$, respectively, the objective functions are formulated as:
	\vspace{-0.1cm}
	\begin{gather}
	\label{eqn:equa1}
	\min_{D} \mathcal{L} \left( D \right) = \mathbb{E}_{\mathbf{\left| X \right|} \in p_{data}} \left [ D \left( \mathbf{\left|X\right|} \right) - 1 \right ]^{2} + \mathbb{E}_{\mathbf{\left| Y \right|} \in p_{\mathbf{z}}} \left [ D \left( G\left(\mathbf{\left|Y\right|} \right) \right)\right ]^{2}, \\
	\min_{G} \mathcal{L} \left( G \right) = \mathbb{E}_{\mathbf{\left| Y \right|} \in p_{\mathbf{z}} }  \left [ D \left( G \left( \mathbf{\left| Y \right|} \right) \right) - 1 \right ]^{2} + \lambda_{G}  \left\| G\left(\mathbf{\left| Y \right|}\right) - \mathbf{\left| X \right|} \right\|_{1},
	\end{gather}
	\vspace{-0cm}
	where $D\left(\cdot \right)$ denotes the probability of data being true. $\lambda_{G}$ denotes the hyper-parameter to weight between the adversarial loss and $\mathcal{L}_{1}$-regularization loss in optimizing G. Note that the regularization term is thought to be necessary and important to recover the speech details~{\cite{pascual2017segan, baby2019sergan}}.
	
	\section{Proposed Network}
	\label{sec:proposed-network}
	This section describes the proposed DARGAN, which is shown in Fig.~{\ref{fig:proposed-architecture}}. First we introduce the generator used herein, then we illustrate the discriminator. Finally we introduce the phase post-processing module.
	
	\subsection{Generator}
	\label{sec:generator}
	In this paper, we use the proposed DARCN~{\cite{li2020recursive}} as the generator module. This is because DARCN has shown satisfactory performance in noise suppression and speech recovery with limited trainable parameters~{\cite{li2020recursive}}. Compared with previous networks~{\cite{wang2018supervised, tan2019learning}}, it combines recursive learning and dynamic attention together. For recursive learning, the training procedure is decomposed into multiple stages. Between adjacent stages, stage recurrent neural network (SRNN) is proposed to bridge the relationship with a memory mechanism~{\cite{li2020recursive}}. Therefore, the estimation in each stage can be refined progressively. To illustrate the point, we formulate the calculation of SRNN at stage $l$ as:
	 
	\begin{equation}
	\label{eqn:equa2}
	\mathbf{h}^{l} = f_{srnn}\left( \left|\mathbf{Y}\right|, \left|\mathbf{\tilde X^{l-1}}\right|, \mathbf{h}^{l-1}   \right)
	\end{equation}
	where $\mathbf{h}$ denotes the state term after SRNN, which is subsequently sent to the afterward module to update the estimation output. $f_{srnn}$ is the mapping function of SRNN. Superscript $l$ refers to the $l$th stage. We can find that the correlation between adjacent stages is bridged through SRNN, which facilitates the estimation afterward.   
	
	Dynamic attention simulates the dynamic property of human auditory perception, i.e., when the real environment changes rapidly, human tends to adjust their auditory attention accordingly. To realize that, a separate attention generator network is designed, which outputs the layer-wise values used to adjust the feature distribution. The coupling method between two modules is via point-wise convolutions and sigmoid functions. To illustrate how the two models operation, we give the calculation process in stage $l$ as:
	
	\begin{gather}
	\label{eqn:equa3}
	\mathbf{a}^{l} = G_{A} \left( \mathbf{|X|}, |\mathbf{\tilde{S}}^{l-1}|\right),\\
	|\mathbf{\tilde{S}}^{l}| = G_{R} \left( \mathbf{|X|}, |\mathbf{\tilde{S}}^{l-1}|, \mathbf{a}^{l} \right),
	\end{gather}
	where $\mathbf{a}$ refers to the values output by generator attention network. $G_{A}$ and $G_{R}$ refer to the mapping function of attention generator network and noise removal network, respectively. 
	
	In this study, the parameter settings of G are the same as~{\cite{li2020recursive}}. As stated in~{\cite{li2020recursive}}, when the number of stages $Q$ equals to 3, the network can adequately balance between performance and computational complexity. So $Q$ is set to 3 in this paper. Additionally, we only apply the supervision to the final stage for computational convenience, which is different from~{\cite{li2020recursive}}. 
	\subsection{Discriminator}
	\label{sec:discriminator}
	In our experiment, we use a typical convolutional recurrent network (CRN) as the discriminator, which is shown in Fig.~{\ref{fig:proposed-architecture}}. It encompasses four parts, namely convolutional-encoder (CE), bidirectional LSTM (BLSTM), fully-connected (FC) layers and adaptive average pool (ADP) layer. For CE, 6 consecutive convolutional blocks are utilized, each of which consists of a convolutional layer, spectral normalization (SN)~{\cite{miyato2018spectral}} and exponential linear unit (ELU)~{\cite{clevert2015fast}}. SN is utilized herein to stabilize the training process of the discriminator. The kernel size and the stride are set to $\left(2, 5 \right)$ and $\left( 1, 2 \right)$ along the temporal and frequency axis, respectively. The number of channels throughout the CE is $\left( 16, 16, 32, 32, 64, 64 \right)$. After the feature encoding, BLSTM is utilized to model the contextual correlations in both directions. Here one BLSTM layer is used, which has 128 units in each direction. After that, we use two FC layers to compress the features and the number of units is $\left(16, 1 \right)$. To tackle the variant length issue of different utterances, the ADP layer is utilized to average the results of all the timesteps, leading to the global result.
	
	\subsection{Phase post-processing}
	\label{sec:phase-post-processing}
	Recently, deep Griffin-Lim algorithm (DGLA) is proposed for phase reconstruction when only clean magnitude is available~{\cite{masuyama2019deep}}, which combines the classical Griffin-Lim algorithm (GLA) and a trainable DNN sub-block together. The diagram is shown in Fig.~{\ref{fig:post-pp}}. In this study, we utilize DGLA as the phase post-processing (PPP) module. Different from using clean magnitude as the reference in~{\cite{masuyama2019deep}}, we provide the estimated magnitude from G as the reference amplitude. Therefore, the spectral phase can be refined by iterating the block multiple times. The block comprises three parts, namely $\mathbf{P_{A}}$, $\mathbf{P_{C}}$ and $\mathbf{\Phi}$, where $\mathbf{P_{A}}$ and  $\mathbf{P_{C}}$ work as the parameter-fixed projection operations, and $\mathbf{\Phi}$ the module with trainable parameters for denoising. We refer the readers to~{\cite{masuyama2019deep}} for details. 
	
%
	
	The calculation procedure within each iteration is given as $\mathbf{ \tilde{X}}^{[m]} = \mathbf{ \tilde{Z}}^{m} - \mathbf{\Phi} \left(\mathbf{\tilde X}^{[m-1]}, \mathbf{\tilde R}^{[m]}, \mathbf{\tilde Z}^{[m]} \right)$. After $M$ iterations, the estimated complex-valued spectrum is denoted as $\mathbf{\tilde X}^{[M]}$. Then the amplitude is normalized to 1 for each T-F bin to extract the phase information. As a consequence, the final estimated complex spectrum after post-processing can be computed as:
	\begin{gather}
	\label{eqn:equa3}
	\mathbf{\tilde X} = \mathbf{A} \odot \mathbf{\tilde X}^{[M]} \oslash \left| \mathbf{\tilde X}^{[M]} \right|,
	\end{gather}
	where $\mathbf{A}$ is the estimated amplitude from GAN, $\odot$ and $\oslash$ denote element-wise multiplication and division, respectively.
	\begin{figure}[t]
		\vspace{-0.3cm}
		\setlength{\abovecaptionskip}{0.235cm}
		\setlength{\belowcaptionskip}{-0.1cm}
		\centering
		\centerline{\includegraphics[width=\columnwidth]{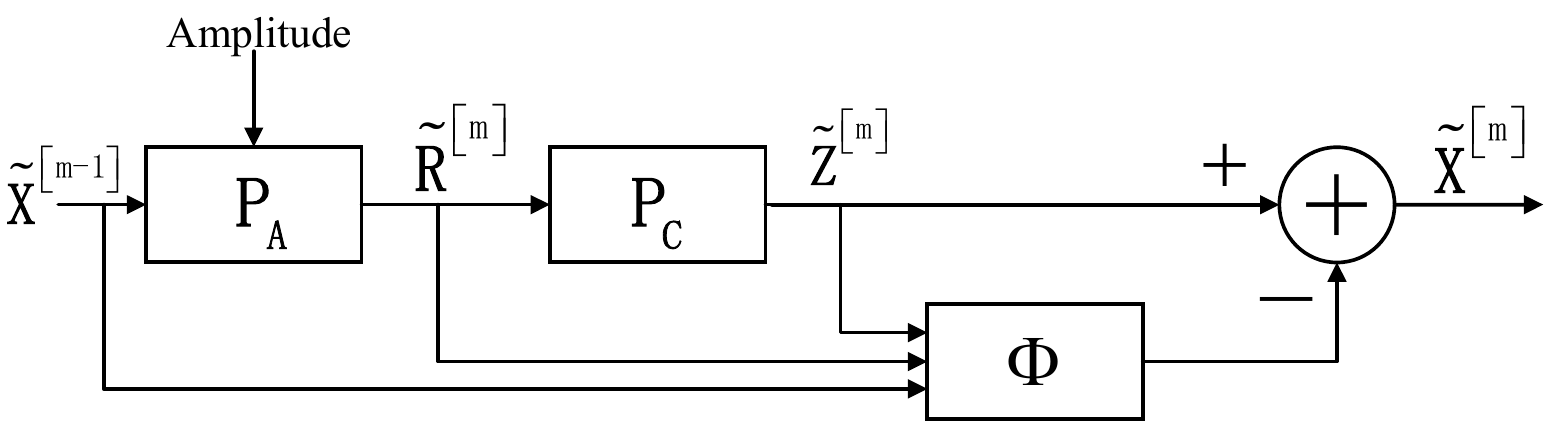}}
		\caption{The schematic of phase post-processing. It is similar to the Deep Griffin-Lim algorithm except the clean magnitude is replaced by the estimated magnitude processed by GAN. $\mathbf{\tilde X}^{[m]}$ is the estimated complex-valued spectrum in the $m^{th}$ iteration.  $\mathbf{\tilde R}$ and $\mathbf{\tilde Z}$ denote the estimated spectrum after $\mathbf{P_{A}}$ and $\mathbf{P_{C}}$, respectively.}
		\label{fig:post-pp}
	\end{figure}
	\vspace{-0.0cm}
	\section{EXPERIMENTAL SETTINGS}
	\label{sec:experimental-settings}
	\subsection{Dataset}
	\label{dataset}
	The experiments are conducted on the dataset released by Valentini et.al.~{\cite{valentini2016investigating}}, which is chosen from the Voice Bank corpus~{\cite{veaux2013voice}}. There are 30 native speakers in total (including 14 male and 14 female), where 28 speakers are used for training (11,572 utterances) and 2 for testing (824 utterances). For each speaker, around 400 utterances are available.
	
	For training noisy set, 10 types of noises are utilized, including 2 synthetic and 8 real from the Demand database~{\cite{thiemann2013diverse}}. 4 SNR levels are set for training: 15\rm{dB}, 10\rm{dB}, 5\rm{dB} and 0\rm{dB}. For testing noisy set, a total of 20 conditions are created: 5 noises from~{\cite{thiemann2013diverse}}  are mixed, each of which is under 4 SNR levels (17.5\rm{dB}, 12.5\rm{dB}, 7.5\rm{dB} and 2.5\rm{dB}). To select the best model during the training, 572 utterances are randomly split from the training set as the validation set. As a result, the number of pairs for training, validation, and testing is 11,000, 572, and 824, respectively. All the utterances are downsampled from 48kHz to 16kHz in our experiment.
	
	\subsection{Network parameter configuration}
	\label{network-configuration}
	The 20ms hamming window is applied, with 50\% overlap between adjacent frames. 320-point STFT is adopted, leading to a 161-D feature vector. For GAN training, the magnitudes of the spectrum are calculated for both the feature and target. We train the network for 100 epochs, optimized by Adam optimizer~{\cite{kingma2014adam}} . The initialized learning rates for G and D are set to 0.0005 and 0.0001, respectively. We only halve the learning rate when consecutive 3 validation loss increment happens and the network is terminated after consecutive 10 validation loss increment arises. The minibatch is set to 4 at the utterance level, the utterance whose length is less than the longest one will be zero-padded.
	
	When the training of GAN is finished, the noisy utterances in the training, validation and testing set are processed with the optimal model, which are then combined with the clean versions to establish new pairs for phase post-processing training. The feature extraction procedure is the same as the previous GAN except the complex-valued spectrum is computed as the feature and the corresponding target. Mean-absolute error (MAE) is used as the training criterion, which is consistent with~{\cite{masuyama2019deep}}. The number of iterations $M$ is set to 5 despite that more iterations can be given. The network is trained for 60 epochs, where the initialized learning rate is set to 0.0002. The minibatch is set to 4 at the utterance level.
	
	\subsection{Baselines}
	\label{baselines}
	To evaluate the performance of the proposed model, a variety of approaches are utilized as the baselines, which are categorized as two types, namely GAN-based methods and non-GAN-based methods. For GAN-based methods, there are SEGAN~{\cite{pascual2017segan}}, SERGAN~{\cite{baby2019sergan}}, GSEGAN~{\cite{fan2019noise}}, MMSE-GAN~{\cite{soni2018time}}, MetricGAN~{\cite{fu2019metricgan}} and CP-GAN~{\cite{liu2020cp}}. For non-GAN-based methods, there are Wavenet~{\cite{rethage2018wavenet}}, Deep Feature Loss (DFL)~{\cite{germain2018speech}}, G+M+P~{\cite{yao2019coarse}}, MDPhD~{\cite{kim2018multi}}, Wave-U-Net~{\cite{giri2019attention}}, WaveCRN~{\cite{hsieh2020wavecrn}} and STFT-TCN~{\cite{koyama2020exploring}}. The reasons for choosing these models are two-fold. First, the metric results of these models are reported on the same dataset~{\cite{valentini2016investigating}}, which makes it possible for fair comparison. Second, the performance of these models are quite competitive.
	
	\subsection{Evaluation Metric}
	\label{evaluation-metric}
	We adopt four metrics to compare the performance of different approaches, which are open-source\footnote{https://www.crcpress.com/downloads/K14513/K14513\_CD\_Files.zip} and described as follows:
	\begin{itemize}
		\item PESQ~{\cite{recommendation2001perceptual}}: Perceptual Evaluation Speech Quality, whose values range from -0.5 to 4.5. The wide-band version is used herein.
		\item CSIG~{\cite{hu2007evaluation}}: Mean opinion score (MOS) prediction related to signal distortion, whose scores range from 1 to 5.
		\item CBAK~{\cite{hu2007evaluation}}: MOS prediction related to background noise, whose scores range from 1 to 5.
		\item COVL~{\cite{hu2007evaluation}}: MOS prediction considering overall quality, whose scores range from 1 to 5. 
	\end{itemize}

	\renewcommand\arraystretch{0.8}
	\begin{table}[t]
		\caption{Experimental results among different models.  We reimplement the results of CSEGAN in~{\cite{fan2019noise}}. N/A denotes the result is not provided in the original paper. PPP denotes the phase post-processing is applied after the GAN estimation. $\textbf{BOLD}$ denotes the best result for each case.}
		\centering
		\LARGE
		\resizebox{0.48\textwidth}{!}{
			\begin{tabular}{cccccc}
				\toprule
				& \multicolumn{1}{c}{Model} & \multicolumn{1}{c}{PESQ}  & \multicolumn{1}{c}{CSIG} & \multicolumn{1}{c}{CBAK} &
				\multicolumn{1}{c}{COVL}\\
				\midrule
				& Noisy &1.97 &3.35 &2.44 &2.63\\
				\midrule
				\multirow{6}*{GAN} 
				& SEGAN~{\cite{pascual2017segan}} &2.16 &3.48 &2.94 &2.80 \\
				& SERGAN~{\cite{baby2019sergan}} &2.62 & N/A & N/A &N/A\\
				& CSEGAN~{\cite{fan2019noise}} &(2.21) &(3.56) &(2.89) &(3.28) \\
				& MMSE-GAN~{\cite{soni2018time}} &2.53 &3.80 &3.12 &3.14 \\
				& MetricGAN~{\cite{fu2019metricgan}} &2.86 &3.99 &3.18 &3.42 \\
				& CP-GAN~{\cite{liu2020cp}} &2.64 &3.93 &3.29 &3.28 \\
				\midrule
				\multirow{6}*{Non-GAN}
				& Wavenet~{\cite{rethage2018wavenet}} &N/A &3.62  & 3.23  & 2.98 \\
				& DFL~{\cite{germain2018speech}} & N/A & 3.86 & 3.33  & 3.22 \\
				& G+M+P~{\cite{yao2019coarse}} & 2.69 &4.00 & 3.34 & 3.34 \\
				& MDPhD~{\cite{kim2018multi}} & 2.70 & 3.85  & 3.39  & 3.27  \\
				& Wave-U-Net~{\cite{giri2019attention}} &2.62 &3.91 &3.32 &3.18 \\
				& WaveCRN~{\cite{hsieh2020wavecrn}} & 2.64 &3.94 &3.37 &3.29 \\ 
				& STFT-TCN~{\cite{koyama2020exploring}} &2.89 &4.24 &3.40 &3.56\\
				\midrule
				\multirow{4}*{Proposed}
				& DARGAN($\lambda_{G}=0.01$) &2.82 &4.22 &3.35 &3.53 \\
				& DARGAN($\lambda_{G}=0.1$) &2.89 &4.23 & 3.39 &3.57 \\
				& DARGAN($\lambda_{G}=1$) &2.93 &$\mathbf{4.30}$ &3.45 &$\mathbf{3.64}$ \\
				& DARGAN($\lambda_{G}=1$) + PPP &$\mathbf{2.96}$ &4.29 &$\mathbf{3.47}$ &$\mathbf{3.64}$ \\
				\bottomrule
			\end{tabular}}
		\label{tbl:results1}
		\vspace{-0.2cm}
	\end{table}
	\vspace{-0.2cm}
	\section{RESULTS AND ANALYSIS}
	\label{sec:results-and-analysis}
	Table~{\ref{tbl:results1}} presents the metric scores of different models. For the proposed network, 3 different values with respect to $\lambda_{G}$ are explored, namely 0.01, 0.1 and 1, indicating different emphasis is given to $\mathcal{L}_{1}$-regularization loss during the training. For $\lambda_{G} = 1$, PPP is also applied to analyze the role of post-processing.
	First, we compare the results of the proposed model with different $\lambda_{G}$. When the value is changed from 0.01 to 1, a notable improvement in all the metrics is achieved. For example, 0.11, 0.08, 0.10 and 0.11 improvement in PESQ, CSIG, CBAK and COVL are observed, which show that a relatively larger weighted coefficient is beneficial to the improvement of speech quality. We have attempted to increase the weighted value but the performance begins to decrease. This is because a further increase in $\lambda_{G}$ will decrease the role of adversarial loss, which may cause implicit damage to the performance.

	We then compare the role of PPP. When PPP is employed as the post-processing technique, slight metric improvement is observed. For example, 0.03 and 0.02 score improvement in terms of PESQ and CBAK is observed, which shows that PPP is beneficial to phase refinement. Note that the performance can be further improved with the increase of the iterations~{\cite{masuyama2019deep}}.

	Finally, we compare the proposed DARGAN with previous works. When comparing previous GAN-based models and DARGAN, we observe a notable improvement. For example, we surpass SEGAN by a large margin in PESQ, CSIG, CBSK and COVL, which are 0.80, 0.81, 0.53 and 0.84, respectively. Even compared with more recently proposed MetricGAN, the proposed model still achieves consistent improvement. When it comes to recently proposed non-GAN methods, the proposed model also gets satisfactory results.  For example, DARGAN outperforms G+M+P, MDPhD, WaveCRN in terms of four objective measurements. We also surpass STFT-TCN, which was used in DNS-Challenge~{\footnote{https://github.com/microsoft/DNS-Challenge}} and ranked fourth in the non-real-time track. It demonstrates the superiority of our proposed model. 

	Fig.~{\ref{fig:visualization}} presents the spectrograms of the utterance enhanced by SEGAN and DARGAN. Form the figure, one can get that the proposed model can effectively suppress the noise components whilst some unnatural residual noise components still remain for SEGAN, as shown in the black dox area of Fig.~{\ref{fig:visualization}}~(c). In addition, compared with SEGAN, the proposed model can also well reserve the speech components like the formant information. More samples are provided at~\url{https://github.com/Andong-Li-speech/DARGAN}.

\begin{figure}[t]
	\centering
	\centerline{\includegraphics[width=0.8\columnwidth]{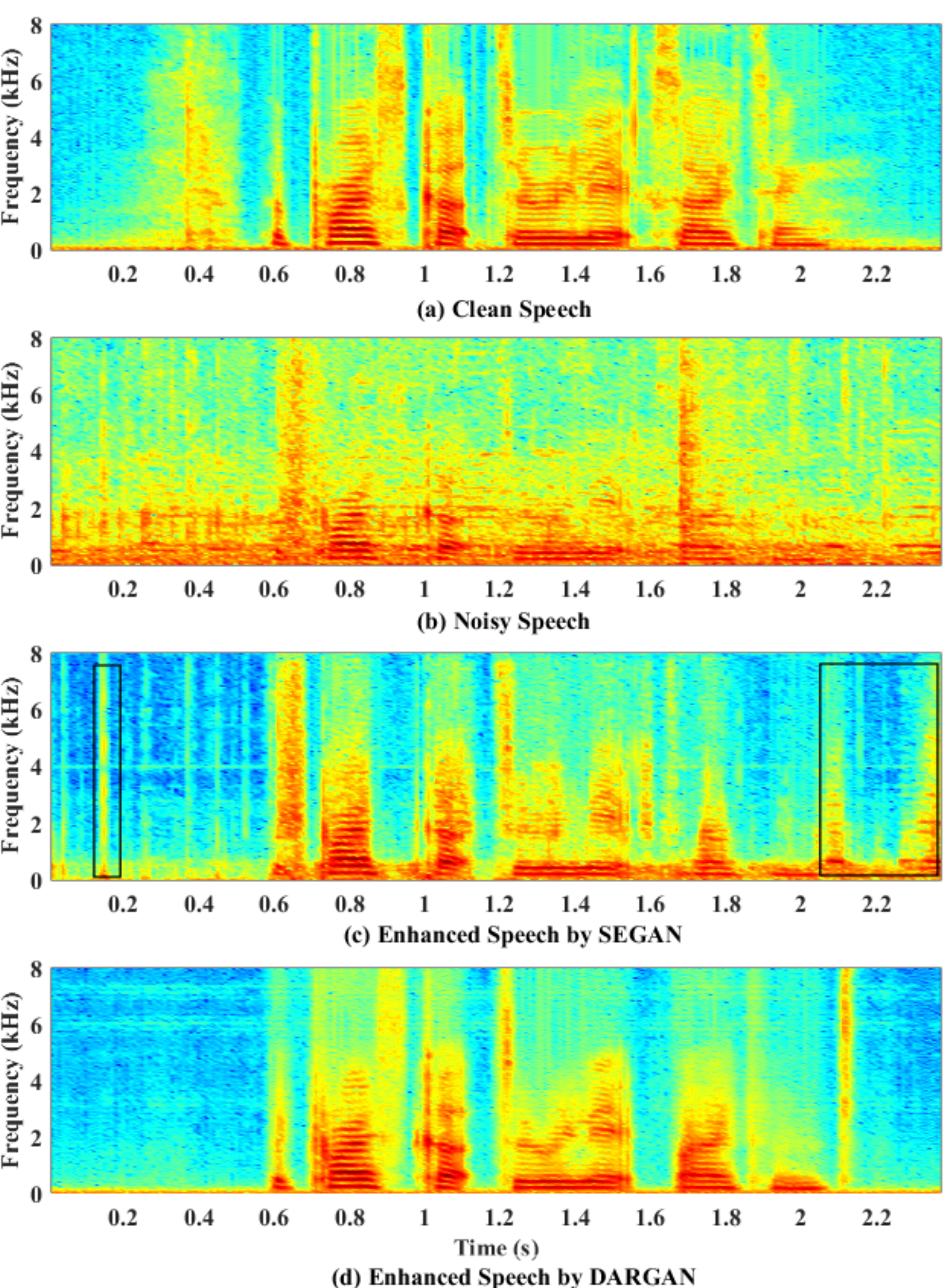}}
	\caption{Visualization of noisy, clean, SEGAN and the proposed model. (a) The spectrum of clean utterance. (b) The spectrum of noisy utterance. (c) The spectrum of the utterance processed by SEGAN. (d) The spectrum of the utterance processed by DARGAN.}
	\label{fig:visualization}
	\vspace{-0.4cm}
\end{figure} 
	
	\vspace{-0.2cm}
	\section{CONCLUSION}
	\label{sec:conclusion}
	\vspace{-0.2cm}
	In this paper, we propose a novel GAN-based model called DARGAN. Compared with previous GAN-based models, three contributions are introduced. First, we adopt the recursive learning, a iterative training protocol to decompose the generating process into multiple stages. Therefore, the estimation result can be refined stage by stage. Second, a dynamic attention mechanism is introduce, where the feature distribution in noise reduction module can be adaptively controlled for better estimation. Third, phase post-processing module is utilized, which facilitates the phase refinement with the increase of module iterations. By doing so, the speech quality can be further improved. Experimental results demonstrate the superiority of DARGAN. Further research involves the direct optimization toward the complex-valued spectrum with GAN.

	\label{sec:refs}
	\small
	\bibliographystyle{IEEEbib}
	\bibliography{refs}

\end{document}